\documentclass[]{article}

 \usepackage[]{graphicx}
 \usepackage{epsfig}
 \usepackage{amsmath, amstext, amsfonts}
 \usepackage{amssymb}
 \usepackage{multicol}
 \usepackage[latin1]{inputenc}

\usepackage{anysize} 
\marginsize{2cm}{2.5cm}{1.5cm}{2cm}

\newcommand{\D}{\displaystyle}

\def\Z{{\mathbb Z}}

\title{Multifractal properties of elementary cellular automata in a discrete wavelet approach of MF-DFA}

\begin{document}

 \maketitle

\begin{center}\sc 
  J.S. MURGUIA$^1$, J.E. P\'EREZ-TERRAZAS$^2$, \& H.C. ROSU$^2$
 \footnotesize \vskip 2ex
       $^1${\em Departamento de F\'{\i}sico Matem\'aticas,
        Universidad Aut\'onoma de San Luis Potos\'{\i},\\
        Alvaro Obreg\'on 64, 78000 San Luis Potos\'{\i}, S.L.P., Mexico\\
       $^2$ IPICYT - Instituto Potosino de Investigaci\'on Cient\'{\i}fica y Tecnol\'ogica,\\
       Camino a la presa San Jos\'e 2055, 78216, San Luis Potos\'{\i}, SLP,
       Mexico}\\
 \end{center} 
 
 \begin{center}

{\tt Europhysics Letters 87 (2009) 28003}\\
celuletas7.tex

\end{center}

\medskip

\noindent PACS {\tt 89.75.Da} - Systems obeying scaling laws\\
PACS {\tt 05.45.Tp} - Time series analysis\\
PACS {\tt 05.40.-a} - Fluctuation phenomena, random processes, noise, and Brownian motion

\bigskip

{
  \footnotesize
   \noindent

   In 2005, Nagler and Claussen (Phys. Rev. E 71, 067103 (2005)) investigated the time series of the elementary cellular automata (ECA) for possible (multi)fractal behavior.
   They eliminated the polynomial background $at^b$ through the direct fitting of the polynomial coefficients $a$ and $b$.
   We here reconsider their work eliminating the polynomial trend by means of the multifractal-based detrended fluctuation analysis (MF-DFA)
   in which the wavelet multiresolution property is employed to filter out the trend in a more speedy way than the direct polynomial fitting and also with respect to the wavelet transform modulus maxima (WTMM) procedure. In the algorithm, the discrete fast wavelet transform is used to calculate the trend as a local feature that enters the so-called details signal. We illustrate our result for three representative ECA rules: 90, 105, and 150.
   We confirm their multifractal behavior and provide our results for the scaling parameters.
   \\

}

\medskip

{\bf Introduction}. -
At the present time, a number of different algorithms are well established to analyse the singular behavior that may be hidden in time series data, such as the WTMM method \cite{muzy-91,muzy2-93,arne1,muzy-94, muzy-95}, the structure function method \cite{muzy2-93}, the DFA \cite{Peng} and its variants \cite{K02,K-review,pawel}. DFA is a method used to analyze the behaviour of the average fluctuations of the data
at different scales after removing the local trends.
In 2002, Kantelhardt {\em et al.} \cite{K02}
provided a generalization of DFA to the case of multifractal time series. Subsequently, the latter method started to be widely employed in the literature under the name of MF-DFA. Kantelhardt wrote a recent review of the techniques used in processing the fractal and MF time series
\cite{K-review}.

On the other hand, as already mentioned, a lot of research has been done on fractal signals and objects with wavelet transforms (WTs) because the multiscale
decompositions implied by the WTs are well adapted to evaluate typical self-similarity properties. The efficiency of WTs as `mathematical microscopes'
for capturing the local scaling properties of fractals have been noticed since more than two decades \cite{argoul}.

\medskip

  It is thus no wonder that there are current efforts towards merging the WTs with DFA procedures \cite{mani} as a natural union of powerful tools for quantifying the scaling properties of the fluctuations.
  In this short note, based on this unifying standpoint, which we call WMF-DFA, we focus on the MF properties of ECA with periodic boundary conditions. There is only one previous work dedicated to the MF features of ECA \cite{jsanchez} but there the analysis is performed on the time series of random walk processes generated by some of the ECA evolution rules and not directly to the ECA time series as we do here. In addition, Nagler and Claussen \cite{nag-05} mention
  in the final part of their work the possibility of considering their spectral analysis for MF signals instead of monofractal ones. We recall that many important applications of ECA are in biology, chemistry, and soft materials, where MF properties are to be expected.
  For example, an interpretation of ECA rules 90 and 150 can be made in the context of catalytic processes \cite{nag-05}, also the rule 126 can be used as a conceptual model of biological cell growth \cite{matache}. On the other hand, the rule 110 is interesting because it has been proven that any mathematical algorithm can be mapped to a ECA having this rule. It is also considered as an intrinsic generator of randomness.

\medskip

 {\bf WMF-DFA}. -
An important advantage of the vanishing moment property of wavelets (see the Appendix) is that it helps detrending the data.
We are interested in revealing the MF properties \cite{Halsey} of ECA. To separate the trend from fluctuations in the ECA time series, we follow the discrete wavelet method proposed by Manimaran {\em et al.} \cite{mani}. This method exploits the fact that the low-pass version resembles the original data in an ``averaged'' manner in different resolutions. Instead of a polynomial fit, we consider the different versions of the low-pass coefficients to calculate the ``local'' trend. Let $x(t_k)$ be a time series type of data, where $t_k=k \Delta t$ and $k=1,2,\ldots,~N$. 
Then the algorithm that we employ contains the following steps (for more mathematical details, see the Appendix):

  \begin{enumerate}

\item  Determine the profile $Y(k)=\sum_{i=1}^{k} (x(t_i)-\langle x\rangle )$ of the time series, which is the cumulative sum of the series from which the series mean value is subtracted.

    \item Compute the fast wavelet transform (FWT), i.e., the multilevel wavelet
          decomposition of the profile.
          For each level $m$, we get the fluctuations
          of the $Y(k)$ by subtracting the ``local''
          trend of the $Y$ data, i.e., $ \Delta Y(k;m) = Y(k) - \tilde{Y}(k;m)$, where $\tilde{Y}(k;m)$ is the reconstructed profile after removal of successive
         details coefficients at each level $m$. These fluctuations at level $m$ are subdivided into windows, i.e., into $M_s={\rm int}(N/s)$ non-overlapping segments of length $s$. This division is performed starting from both the beginning and the end of the fluctuations series (i.e., one has $2M_s$ segments). Next, one calculates the local variances associated to each window $\nu$ 
         \begin{equation}\label{eq-Fs1}
           F^2(\nu,s;m)={\rm var}\Delta Y((\nu-1)s+j;m)~, \quad j=1,..., s~, \quad \nu=1,..., 2M_s~, \quad M_s={\rm int}(N/s)~.
         \end{equation}

    \item  Calculate
           a
           $q$th order fluctuation function defined
           as

          \begin{equation}\label{eq-Fqs}
            F_q(s;m) = \left\{ \frac{1}{2M_s} \sum_{\nu=1}^{2M_s} |F^2(\nu,s;m)|^{q/2} \right\}^{1/q}
          \end{equation}

          where $q \in \Z$ with $q \neq 0$. Because of the
          diverging exponent when $q\to 0$ we employed in this limit
          a logarithmic averaging $\D F_0(s;m) = \exp\left\{\frac{1}{2M_s} \sum_{\nu=1}^{2M_s} \ln |F^2(\nu,s;m)| 
          \right\}$ as in \cite{K02, telesca}.
  \end{enumerate}

    In order to determine if the analysed time series have a fractal scaling behavior, the fluctuation function $F_q(s;m)$
    should reveal a power law scaling
       \begin{equation}\label{eq-FqsPLaw}
         F_q(s;m) \sim s^{h(q)},
       \end{equation}
   where $h(q)$ is called the generalized Hurst exponent \cite{telesca} since it can depend on $q$, while the original Hurst exponent is $h(2)$.
   If $h$ is constant for all $q$ then the time series is monofractal, otherwise it has a MF behavior. In the latter case, one can calculate various other MF scaling exponents, such as $\tau(q)$ and $f(\alpha)$ \cite{Halsey}.

\medskip

 {\bf Application to ECA}. -
We apply the previous algorithm to the time series of three illustrative ECA as classified by Wolfram in 1984 \cite{W84}. The chosen rules are the following: 90, 105, and 150. For the first and the last rules the updates are given by
\begin{equation}\label{eca-1}
x_n^{t+1}=[x^{t}_{n-1}+rx^{t}_{n}+x^{t}_{n+1}]{\rm mod}\, 2~,
\end{equation}
where $r=0$ and $r=1$, respectively.
It is well known that rule 90 has the appearance of a Sierpinski triangle when responding to an impulse (first row is all 0s with a 1 in the center).
Nagler and Claussen \cite{nag-05} found that the rule 150 displays a Sierpinski-like self-similar structure of fractal dimension $d_F=1.69$ (golden mean) instead of the standard one of 1.58. In a subsequent paper \cite{claus08}, Claussen showed that its time behavior can be solved as a two-step vectorial, or string, iteration, which can be viewed as a generalization of Fibonacci iteration generating the time series from a sequence of vectors of increasing length. This could explain the difference in the fractal dimension.
As for the rule 105, it is known to be complementary to the rule 150, i.e., $f_{105}=1-f_{150}$, where $f$ is the neighborhood-depending updating rule.

\begin{figure}[h!]
 \centering
  \includegraphics[width=11.5cm, height=12cm]  {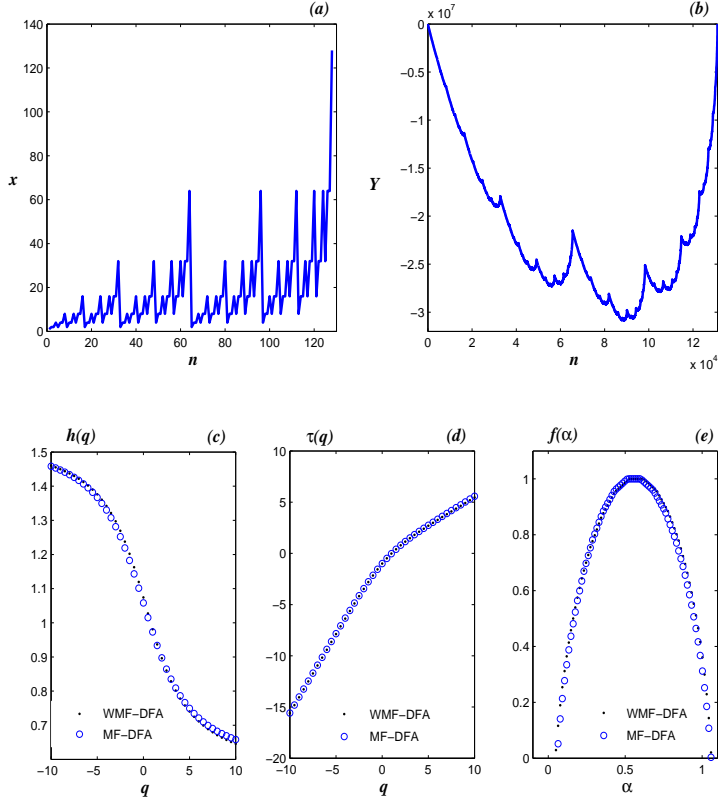}
   \caption{\sl \small
   Rule 90: (a) Time series of the row signal. Only the first $2^7$ points are shown of the whole set of $2^{17}$ data points. (b) Profile $Y$ of the row signal. (c) Generalized Hurst exponent $h(q)$. (d) The $\tau$ exponent, $\tau(q)=qh(q)-1$. (e) The singularity spectrum $f(\alpha)=q\frac{d\tau(q)}{dq}-\tau(q)$.
   The calculations of the multifractal quantities $h$, $\tau$, and $f(\alpha)$ are performed both with MF-DFA and the wavelet-based WMF-DFA.
  }
  \label{fig-90}
\end{figure}

\begin{figure}[h!]
 \centering
  \includegraphics[width=11.5cm, height=12cm]  {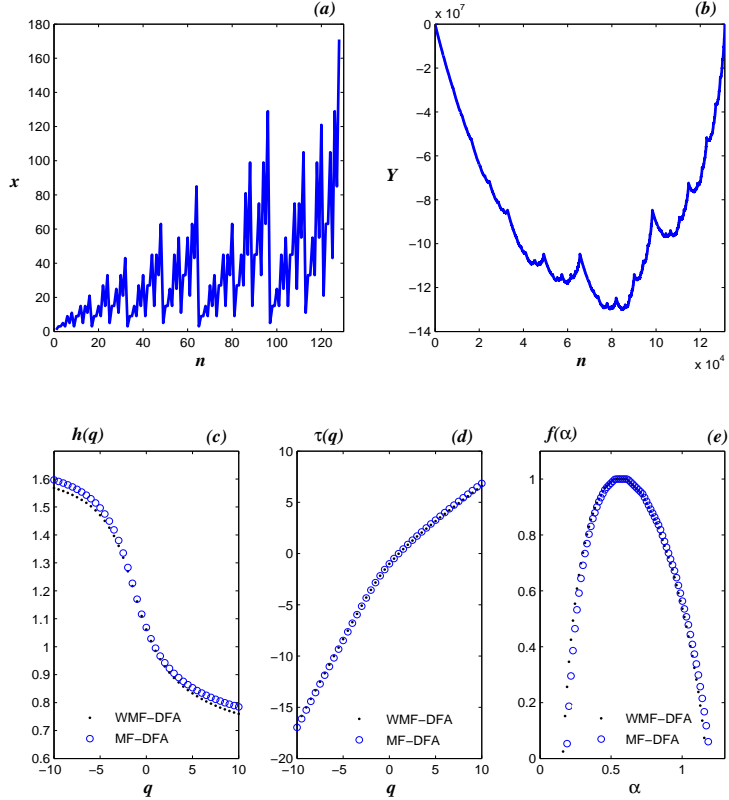}
   \caption{\sl \small
   Same plots as in Fig.~\ref{fig-90} but for rule 150.
  }
  \label{fig-150}
\end{figure}

\begin{figure}[h!]
 \centering
  \includegraphics[width=11.5cm, height=12cm]  {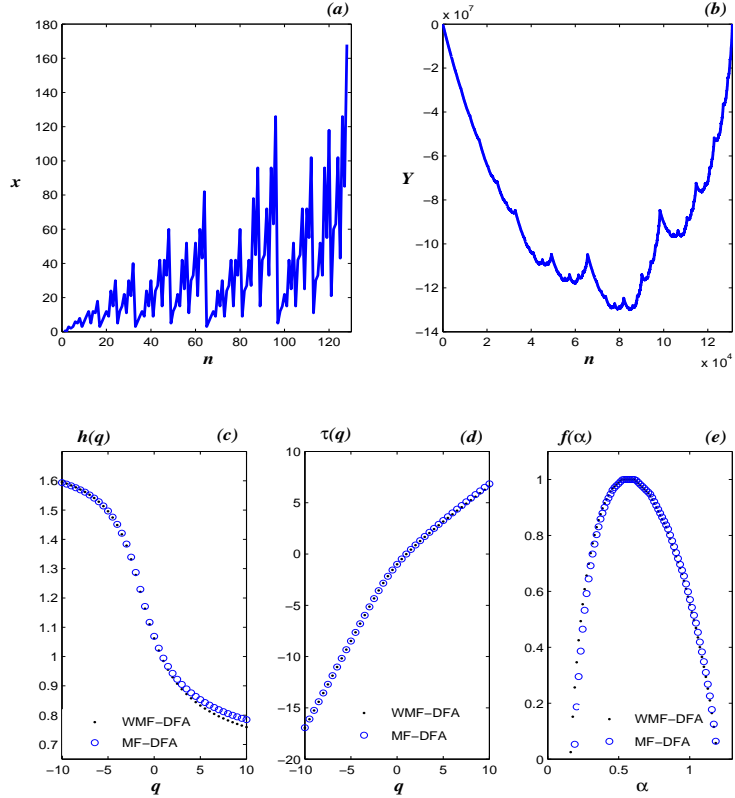}
   \caption{\sl \small
   Same plots as in the previous figures but for rule 105.
  }
  \label{fig-105}
\end{figure}

\begin{figure}[h!]
 \centering
  \includegraphics[width=9.5cm, height=14.5cm]  {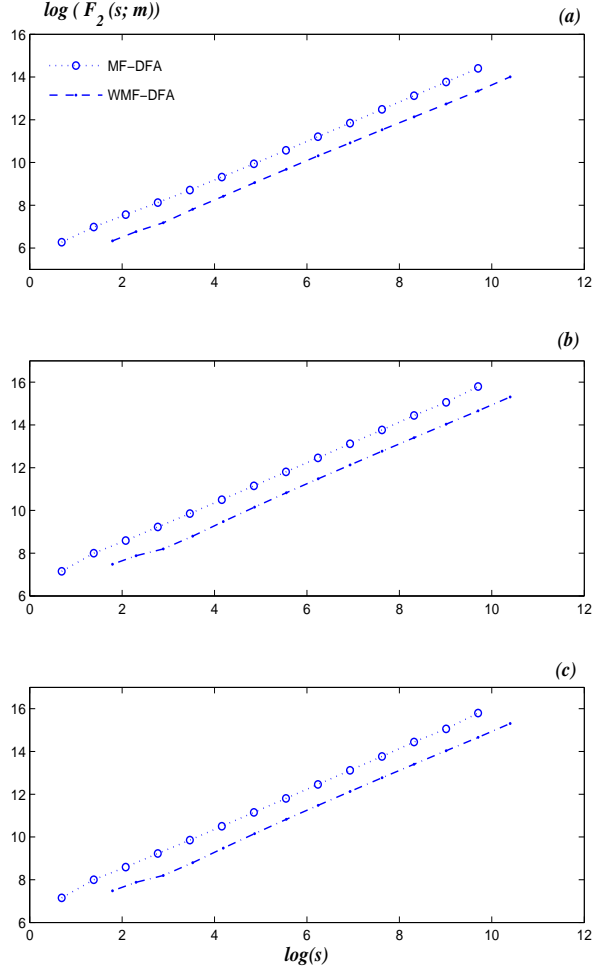}
   \caption{\sl \small
   Log-log plot of the fluctuation function $F_2$ versus scale for: (a) rule 90, (b) rule 150, and (c) rule 105.
  }
  \label{fig-dfa}
\end{figure}

\begin{table}[h!]
     \caption{The Hurst exponent $h(2)$ for the three ECA rules examined in this paper as obtained by means of the MF-DFA and WMF-DFA methods and in each of the cases for four initial (first row) pulses as indicated.}
      \label{T-s1-mat}
     {\small
     \begin{center}
     \begin{tabular}{|r|l|l|l|l|l|l|l|l|}
        \hline \hline
        & \multicolumn{4}{|c|}{MF-DFA} & \multicolumn{4}{|c|}{WMF-DFA} \\
        \hline \hline
        & I  & II  &   III  &  IV &      I  &     II &    III &  IV  \\
        & \tiny{($\cdots 010 \cdots)$} & \tiny{($\cdots 0110 \cdots)$} & \tiny{($\cdots 01010 \cdots)$} & \tiny{($\cdots 01110 \cdots)$}
        & \tiny{($\cdots 010 \cdots)$} & \tiny{($\cdots 0110 \cdots)$} & \tiny{($\cdots 01010 \cdots)$} & \tiny{($\cdots 01110 \cdots)$}\\
        \hline \hline
   R90  & 0.8972 & 0.8972 & 0.8898 & 0.9451 & 0.8961 & 0.8961 & 0.9229 & 0.9787 \\
   R150 & 0.9427 & 0.9413 & 0.9541 & 0.9296 & 0.9293 & 0.9529 & 0.9513 & 0.9407 \\
   R105 & 0.9427 & 0.9413 & 0.9542 & 0.9296 & 0.9294 & 0.9529 & 0.9514 & 0.9407 \\
   \hline
    \end{tabular}
     \end{center}
     }
     \end{table}

\bigskip

We have analysed the time series of the so-called row sum (or total activity) ECA signals, i.e., the sum of ones in sequences of rows, employing Daubechies (Db) wavelets. Various types of Db wavelets have been used but we have found that a better matching of the results given by the WMF-DFA method with those of other methods is provided by the Db-4 wavelets with four filter coefficients. Our results are illustrated in Figs.~(\ref{fig-90})-(\ref{fig-105}). The fact that the generalized Hurst exponent is not a constant horizontal line is indicative of a multifractal behavior in all three cases. In addition, the fact that the $\tau$ index is not of a single slope is another clear feature of multifractality.
The values of the Hurst exponent $h(2)$ for four types of initial conditions are given in Table~1. We also present the corresponding fluctuation function $F_2$ in Fig.~(\ref{fig-dfa}) for the impulsive initial condition.
The strength of the multifractality is roughly measured with the width $\Delta \alpha = \alpha _{\rm max}- \alpha_{\rm min}$ of the parabolic singularity spectrum $f(\alpha)$ on the $\alpha$ axis.
For example, for the impulsive initial condition, $\Delta \alpha  _{90}=0.9998 (1.0132)$, $\Delta \alpha  _{150}=1.011 (1.0075)$, and $\Delta \alpha  _{105}=1.0083 (1.0325)$ when the MF-DFA (WMF-DFA) are employed.
We notice that the most ``frequent'' singularity for all the analysed time series occurs at $\alpha = 0.568$, where the width $\Delta \alpha$
of rule 90 is shifted to the right with respect to those of 105 and 150. According to our results, the strongest singularity, $\alpha_{\rm{min}}$, of all time series corresponds to the rule 90 and the weakest singularity, $\alpha_{\rm{max}}$, to the rule 150.

In conclusion, in general terms, our algorithm implementation shows that embedding the discrete wavelet transform in the MF-DFA technique is a well-suited procedure to analyze the multifractal properties of the ECA. Indeed, we get similar results to the other methods but computationally faster because we employ a lesser number of windows. Our results represent a confirmation of the fact that ECA patterns of different magnitudes follow different scaling laws, i.e., the ECA have intrinsic multifractality that does not depend on the set of initial data that we used. Therefore, when processes thought to be multifractal are simulated with (E)CA, their intrinsic multifractal behavior should be taken into account as a feature of the simulation procedure rather than of the multifractal behavior of the simulated processes.

\newpage

\section*{Appendix}

{\em WT: continuous and discrete}. -
The WT of a function or distribution function $x(t)$ is given by
    \begin{equation}
      W_x(a, b) = \frac{1}{a} \int_{-\infty}^{\infty} x(t)\bar{\psi} \left(\frac{t - b}{a} \right) dt,
        \label{eq-CWT}
   \end{equation}
    where $\psi$ is the analyzing wavelet, $b \in \mathbb{R}$ is a translation parameter, whereas $a \in \mathbb{R}^{+} ~ (a \neq 0)$ is a dilation or scale parameter, and the bar symbol denotes complex conjugation.
    One fundamental property that we require in order to analyze the singular behavior of a signal is that $\psi(t)$ has enough vanishing moments~\cite{arne1, mallat}. A wavelet has $n$ vanishing moments if and only if it satisfies $\int_{-\infty}^{\infty} t^k \psi(t) dt  = 0$ for
    $k = 0, 1, \ldots , n - 1$ and $\int_{-\infty}^{\infty} t^k \psi(t) dt  \neq 0$ for $k = n$.
This means that a wavelet with $n$ vanishing moments is orthogonal to all polynomials up to order $n-1$. Thus, the WT of $x(t)$ performed with a wavelet $\psi(t)$ with $n$ vanishing moments is nothing else but a ``smoothed version'' of the $n$'th derivative of $x(t)$ on various scales.

    \medskip

    Since the ECA data are notoriously discrete it is important to consider a
    discrete version of \eqref{eq-CWT}. Generally,
    the orthogonal (discrete) wavelet transform (DWT) is employed. This is only one of
    the different forms of WTs \cite{Mallat}, by which
    the wavelets are associated to orthonormal
    bases of $L^2(\mathbb{R})$. In this case, the wavelet transform is performed
    only on a discrete grid of the parameters of dilation and translation, i.e., $a$ and $b$ take on only integer
    values. In fact, for the numerical implementation of the DWT the multiresolution
    analysis (MRA) has been introduced.

   The representation of a function or process $x(t)$ with the DWT is given in terms of shifted and dilated
   versions of the wavelet function $\psi(t)$,
   and its associated scaling function $\varphi(t)$
   \cite{Mallat, Ingrid}. Within this framework  and
   considering that the scaling and wavelet functions
 \begin{equation}
   \label{eq-TO-base}
   \varphi_{m,n}(t) = 2^{m/2} \varphi(2^m t - n), \qquad
   \psi_{m,n}(t) = 2^{m/2} \psi(2^m t - n), \quad m,n\in \Z
 \end{equation}
 %
form an orthonormal basis, then one can write the expansion of $x(t)$ 
as follows
  \begin{equation}
         \label{eq-TO-AMR}
     x(t) = \sum_n \left(a_{m_0,n} \varphi_{m_0,n}(t)+ 
      \sum_{m = m_0}^{M-1}d_{m,n} \psi_{m,n}(t)\right),
  \end{equation}
where the scaling or approximation coefficients $a_{m,n}$, and the wavelet coefficients $d_{m,n}$ are defined as
  \begin{equation}
         \label{eq-coef-c-d}
     a_{m,n} = \int x(t) \varphi_{m,n}(t) dt, \qquad
     d_{m,n} = \int x(t) \psi_{m,n}(t) dt,
  \end{equation}
 with $m$ and $n$ denoting the dilation and translation indices,
   respectively.\\

   \medskip

   {\em FWT}. -
    To calculate $ a_{m,n}$ and $d_{m,n}$, Mallat \cite{Mallat} developed the FWT in which the MRA approach is involved.
    The FWT algorithm connects, in an elegant way, wavelets and filter banks, where the multiresolution signal decomposition of a signal $X$, based on successive decomposition, is composed by a series of approximations and details which become increasingly coarse.
    At the beginning, the signal is split into an approximation and a detail part that together yield the
    original. The subdivision is such that the approximation signal contains the low frequencies, while the detail
    signal collects the remaining high frequencies. By repeated application of this subdivision rule on the approximation, details of increasingly
    coarse resolution are separated out, while the approximation itself grows coarser and coarser.

    The FWT calculates the scaling and wavelet coefficients at scale $m$
    from the scaling coefficients at the next finer scale $m + 1$ using the following formulas

      \begin{align}
        \label{eq-TO-pro}
      a_{m,n} & = \sum_k h[k - 2n] a_{m+1,k}, \\
        \label{eq-TO-det}
      d_{m,n} & = \sum_k g[k - 2n] a_{m+1,k},
     \end{align}
   %
 where $h[n]$ and $g[n]$ are typically called low pass and high pass filters in the associated analysis filter bank.
In fact, the signals $a_{m,n}$ and $d_{m,n}$ are the convolutions of $a_{m+1,n}$ with the filters $h[n]$ and $g[n]$ followed by a downsampling
of factor 2 \cite{Mallat}, respectively.

\begin{figure}[h!]
 \centering
  \includegraphics[width=7.5cm, height=5cm]{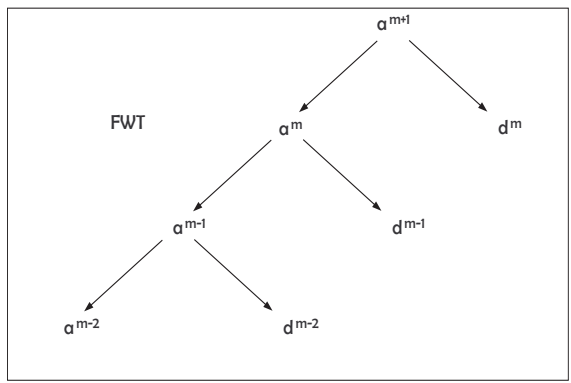}
   \caption{\sl \small
    The structure of a three-level FWT.
  }
   \label{fig-FWT}
\end{figure}

Conversely, a reconstruction of the original scaling coefficients $a_{m + 1,n}$ can be made from the following combination of the scaling and wavelet coefficients at a coarse scale
\begin{equation} \label{eq-TO-FBS}
      a_{m + 1,n} = \sum_k \left( h[2k - n] a_{m,k} + g[2k - n] d_{m,k} \right)~.
   \end{equation}
   %
It corresponds to the synthesis filter bank. This part can be viewed as the discrete convolutions between the upsampled signal $a_{m,l}$ and the filters $h[n]$ and $g[n]$, that is, following an upsampling of factor 2 the convolutions between the upsampled signal and the filters
$h[n]$ and $g[n]$ are calculated. The number of levels depends on the length of the signal, i.e., a signal with $2^L$ values can be decomposed into $(L+1)$ levels.
    To initialize the FWT, we consider a discrete time signal $X =\{x[1], x[2], \ldots, x[N]\}$ of length $N =2^L$. The first application of \eqref{eq-TO-pro} and \eqref{eq-TO-det}, beginning with $a_{m+1,n}=x[n]$, defines the first level of the FWT of $X$.
    The process goes on, always adopting the $m+1$th scaling coefficients to calculate the ``$m$'' scaling and wavelet coefficients. Iterating \eqref{eq-TO-pro} and \eqref{eq-TO-det} $M$ times, the transformed signal consists of $M$ sets of wavelet coefficients at scales $m=1, \ldots, M$, and a signal set of scaling coefficients at scale $M$. There are exactly $2^{(L-m)}$ wavelet coefficients $d_{m,n}$ at each
    scale $m$, and $2^{(L-M)}$ scaling coefficients $a_{M,n}$. The maximum number of iterations is $M_{\max}=L$. A three-level decomposition process
    of the FWT is shown in Figure \ref{fig-FWT}.

\begin{center} *** \end{center}

 JSM received partial financial support from PROMEP and FAI--UASLP,
 JEPT received an IPICyT student fellowship, and HCR got partial support through a CONACyT project.



\begin{thebibliography}{50}

\bibitem{muzy-91} MUZY J.F., BACRY E. and ARNEODO A., 
                  {\em Phys. Rev. Lett.}, {\bf 67} (1991) 3515.


 \bibitem{muzy2-93}
      MUZY J.F., BACRY E. and ARNEODO A.,
      {\em Phys. Rev. E}, {\bf 47} (1993) 875.



  \bibitem{arne1}
          BACRY E., MUZY J.F. and ARNEODO A.,
          {\em J. Stat. Phys.}, {\bf 70} (1993) 635.


  \bibitem{muzy-94}
     MUZY J.F., BACRY E. and ARNEODO A.,
     {\em Int. J. Bif. and Chaos}, {\bf 4} (1994) 245.

  \bibitem{muzy-95}
     ARNEODO A., BACRY E. and MUZY J.F.,
     {\em Physica A}, {\bf 213} (1995) 232.

        \bibitem{Peng}
     PENG C.-K., BULDYREV S.V., HAVLIN S., SIMONS M., STANLEY H.E. and GOLDBERGER A.L.,
     {\em Phys. Rev. E}, {\bf 49} (1994) 1685.

   \bibitem{K02}
     KANTELHARDT J.W., ZSCHINEGNER S.A., KOSCIELNY-BUNDE E., HAVLIN S., BUNDE A. and STANLEY H.E., 
     {\em Physica A}, {\bf 316}, (2002) 87.

     \bibitem{K-review} KANTELHARDT J.W., 
     {\em Springer Encyclopedia of Complexity and System Science}, edited by MEYERS R.A. (Springer Science+Business Media, LLC., New York) 2009.


   \bibitem{pawel}
           OSWIECIMKA P., KWAPIEN J. and DROZDZ S.,
           {\em Phys. Rev. E}, {\bf 74} (2006) 016103.

\bibitem{argoul} ARGOUL F., ARNEODO A., ELEZGARAY J., GRASSEAU G. and MURENZI R.,
{\em Phys. Lett. A}, {\bf 135} (1989) 327.

  \bibitem{mani}
   MANIMARAN P., PANIGRAHI P.K. and PARIKH J.C.,
   {\em Phys. Rev. E}, {\bf 72} (2005) 046120.


    \bibitem{jsanchez}
    SANCHEZ J.R.,
        {\em Int. J. Mod. Phys. C}, {\bf 14} (2003) 491.

   \bibitem{nag-05}
   NAGLER J. and CLAUSSEN J.C.,
   {\em Phys. Rev. E}, {\bf 71} (2005) 067103.

   \bibitem{matache} MATACHE M.T. and HEIDEL J.,
   {\em Phys. Rev. E}, {\bf 71} (2005) 026232.

   \bibitem{Halsey}
    HALSEY T.C., JENSEN M.H., KADANOFF L.P., PROCACCIA I. and SHRAIMAN B.I.,
    {\em Phys. Rev. A}, {\bf 33} (1986) 1141.


  \bibitem{telesca}
        TELESCA L., COLANGELO G., LAPENNA V. and MACCHIATO M.,
        {\em Phys. Lett. A}, {\bf 332} (2004) 398.

        \bibitem{W84} WOLFRAM S., {\em Physica D}, {\bf 10} (1984) 1.

        \bibitem{claus08} CLAUSSEN J.C.,
        {\em J. Math. Phys.}, {\bf 49} (2008) 062701.

  \bibitem{mallat}
      MALLAT S. and HWANG W.L.,
      {\em IEEE Trans. Inform. Theory}, {\bf 38}(2) (1992) 617.


\bibitem{Mallat}
    MALLAT S., {\it A Wavelet Tour of Signal Processing}
    (Academic Press, New York) 1999.

   \bibitem{Ingrid}
     DAUBECHIES I., {\it Ten Lectures on Wavelets} (SIAM, Philadelphia,
     Penn.) 1992.


\end{thebibliography}
\end{document}